# CITIZEN SCIENCE IN THE EUROPEAN OPEN SCIENCE CLOUD

■ **Stephen Serjeant** – The Open University – **DOI:** https://doi.org/10.1051/epn/2023204

In 2007, Michael Longo published a startling result: he appeared to have found a fundamental handedness to the large-scale structure of the Universe, in an asymmetry between clockwise and anticlockwise spiral galaxies. His sample of 22 thousand galaxies had been examined partly by eye and partly algorithmically, and so the challenge was laid down to the community to verify or refute this result.

▲ A selection of letter-like galaxies spotted by volunteers in the Galaxy Zoo citizen science experiment. Created with mygalaxies.co.uk

**B**ut eyeballing 22 thousand galaxies, or more, is no small task. I had it on my own whiteboard on a long to-do list of things that ought to be done by someone and that it would be nice to do if I had time, which of course meant never. But the team behind Galaxy Zoo hit on what turned out to be a genuinely brilliant solution: crowdsource the job with the help of volunteers from the general public.

And it was a runaway success. In as little as three weeks, the volunteers classified a million galaxies, with each galaxy being checked at least 20 times. It would have taken a postdoc five years. The evidence for spiral handedness went away, but Galaxy Zoo team hit on a very rich seam: people genuinely want to participate in scientific discovery and are willing to spend their own time doing it. Galaxy Zoo has inspired well over a hundred crowdsourced data mining citizen science projects, hosted on the Zooniverse platform.

This is now a well told story, but there's one message that's often forgotten: from the outset, this citizen science platform has always been intended as a tool for doing research. It's a biological computer, a multi-headed hydra of a hundred thousand human brains or more, all deployed to solve a research goal. In this sense it's no different to a spectrometer, or an accelerator beamline. And as with any other scientific tool or facility, there are science questions that are particularly well suited to the tool, and ones that are not.

What this citizen science is not primarily designed for is outreach or education – which is not to say that it can't inspire or guide volunteers into deeper engagement with science, but if your main goal is to inspire or to educate, then my consistent advice is not to do citizen science. Rather, spend your efforts directly on inspirational public engagement and education. But if you have a complex data mining problem, then citizen science should be among your tools. It's a missed opportunity, and a fundamental misunderstanding, when citizen science is said to be synonymous with outreach.

Meanwhile, during the burgeoning successes of citizen science, there has also been a simultaneous growth of open science initiatives. The European Commission in particular has earmarked well over a quarter of a billion Euros on the development and deployment of the European Open Science Cloud, or EOSC. The citizen community is one of the strategic priorities of the Strategic Research and Innovation Agenda of the EOSC (https://eosc.eu/sria). Even so, only a very small fraction of the resource is being deployed on their engagement with the European Open Science Cloud.

Citizen science nevertheless involves an enormously larger and more diverse scientific user community with the European Open Science Cloud. For example, our Galaxy Zoo Clump Scout project had a science team of just three academics but a community of nearly fourteen thousand volunteers, contributing nearly two million classifications. In the Clump Scout project, as with all our citizen science projects, contextual educational and training resources are embedded into the volunteer workflows. This allows non-specialist volunteers to gain enough subject specialist knowledge for more comprehensive explorations of the data, and indeed on many projects there are explicit links to external professional tools for this deeper engagement. There is also evidence for volunteers acquiring new scientific terminology that was not provided in training material, *i.e.* there is evidence that the activity has stimulated independent study (*e.g.* Luczak-Roesch *et al.* 2014, *International AAAI Conference on Web and Social Media*; Oesterlund *et al.* 2017, *Academy of Management Annual Meeting Proceedings*). Nevertheless, the education is part of the supporting structure and is not the primary goal; it is always exclusively in the context of volunteers participating in projects with clear science goals, which drive and define the data mining activities.

A central vision of the European Open Science Cloud is to make scientific data FAIR, that is Findable, Accessible, Interoperable and Reusable. Implicit in this vision is that the FAIR data should also be useful, but this is far from being guaranteed, especially given its inter-disciplinary and multi-disciplinary remit. For example, Daylan *et al.* 2016 (*Physics of the Dark Universe*, Volume 12, p. 1-23) reanalysed public sky survey data from the Fermi gamma-ray telescope, and interpreted a gamma-ray excess ● ● ●

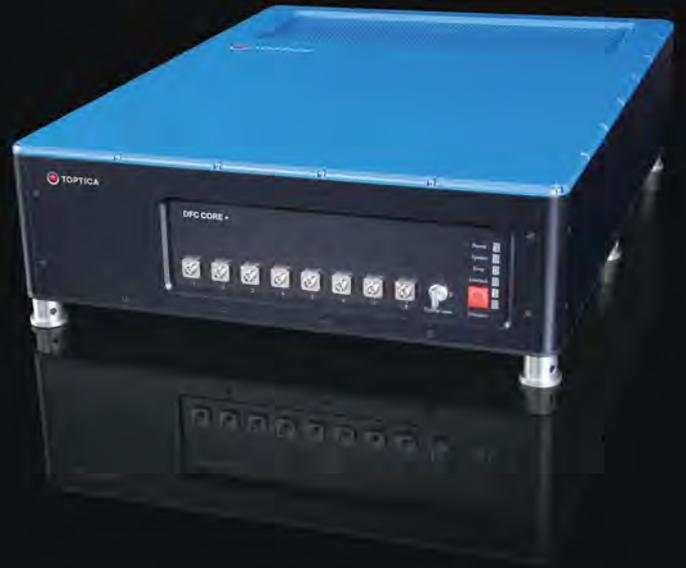
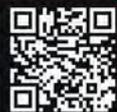
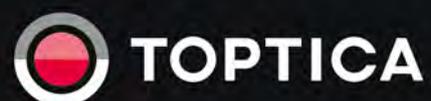

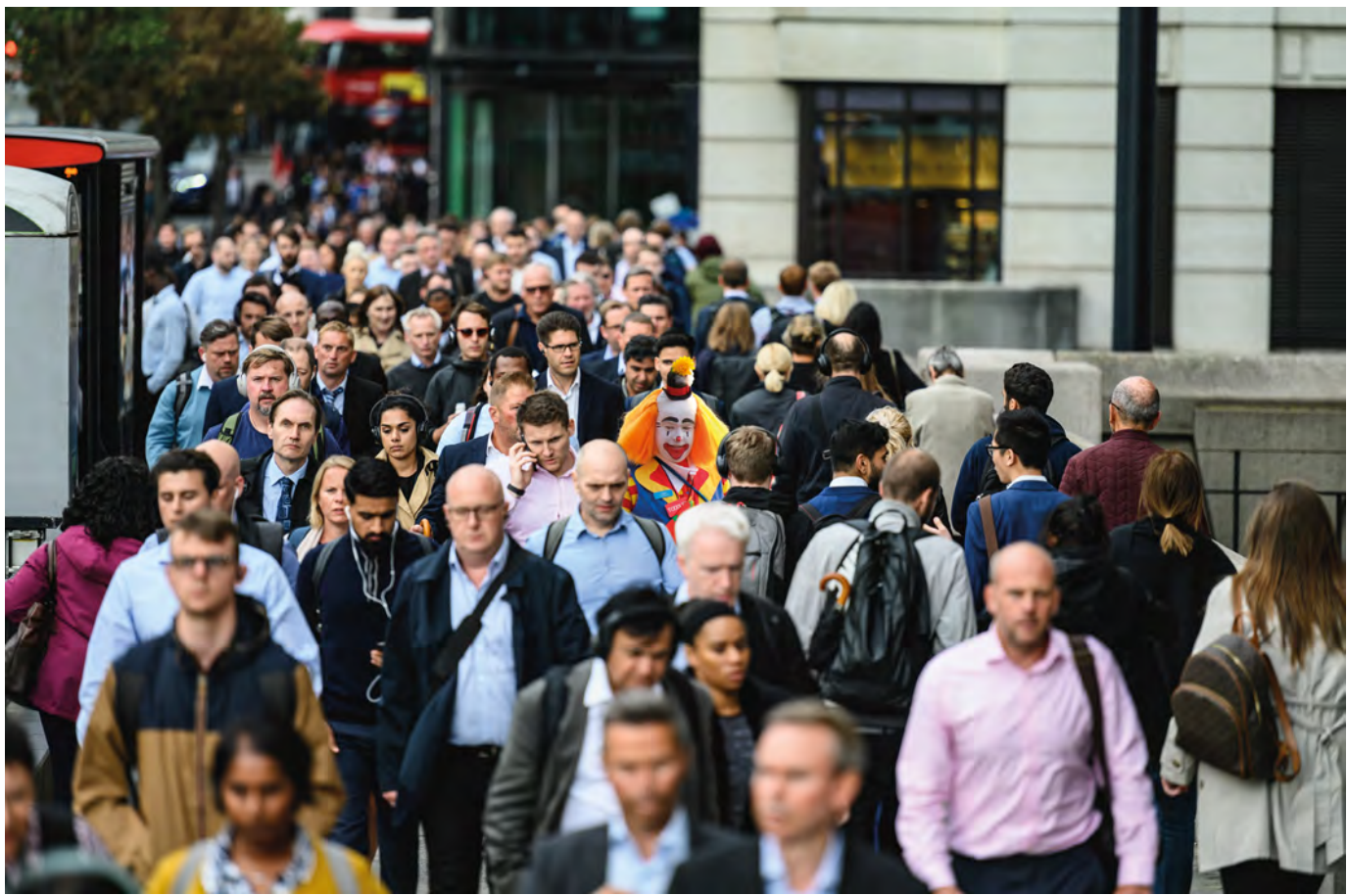

◂ **FIG. 1:**
A facial recognition algorithm would tell you where the faces are, but at the moment only a human has the ability to "un-ask" the question and say, wait a minute, there's a clown in the image!

towards the Galactic centre as a signature of dark matter particle annihilation. This would be the natural location for such a signal, given the expected density peak in the dark matter distribution. However, the instrument team themselves responded to this claim (Ackermann *et al.* 2017, *Astrophysical Journal* **840-1**, article id. 43) by re-interpreting the claimed signal as an observational systematic. Without taking any view on the merits or otherwise of either side of this debate, it's clear that the usefulness of FAIR data will be limited by the supporting supplementary contextual information, beyond metadata even into training. The further that the data are from a user's subject specialism, the more curated the interaction must be with that data. The most extreme example of this is citizen science. In astronomy, there aren't many serious consequences for misunderstanding and misuse of data, but in *e.g.* healthcare or climate science the stakes are much higher and the specialist communities have to take much more care.

But in astronomy at least, we are safer to experiment with bringing the public into open science. And so, my Open University colleagues Hugh Dickinson and James Pearson have been building Jupyter notebooks that demonstrate how to design, build and run citizen science projects, as well as fold in machine learning such as the excellent deep learning tools by Mike Walmsley. The goal is to have plug-and-play exemplars to help the professional community engage with both the EOSC open science tools and with the abundant research effort available from the volunteer community.

Our vision for a professional-amateur collaboration would work like this, at least in the context of the EOSC products and services for astronomy and astroparticle physics built by the EOSC ESCAPE project (European Science Cluster of Astronomy & Particle Physics ESFRI Research Infrastructures, https://projectescape.eu). We start with the professional scientists working already on the ESCAPE science analysis platform, using data from its data lake. On that platform they are building and running data mining projects, initially deploying volunteers in citizen science. In many cases in astronomy, citizen science volunteers are already able to link out of the citizen science projects into the many accessible professional EOSC virtual observatory tools, armed with the new knowledge that they have acquired as part of the citizen science project that gives them the context of the scientific data. The professionals collate and (where necessary) reduce or clean the volunteer data on the science analysis platform, then use it as a training set for machine learning to accelerate the classifications. One can then use the machines to classify the most straightforward and unambiguous items to classify, and refocus the human effort on the difficult edge cases that are the most sensible use of human effort. This drives a virtuous circle between human and machine classification, but the human effort still maintains advantages over machines, not just in being able to identify or classify items that are too rare to train machine learning for, but also in being able to take a step back and "un-ask" the fundamental question or task (Figure 1).

As with any scientific tool, there is a craft to doing citizen science well. For a project to be successful the classification task often needs to be visually attractive, and the research must be manifestly important – because after all you're asking a hundred thousand people to work for you, so it had better be worthwhile. Often, the simpler volunteer workflows get more engagement. Most importantly, the academic team need to be prepared to put in the time to engage in online forums with their volunteers, and they will need to clean the data when it arrives. It's also worth keeping in mind what the volunteer rewards and motivations could be. Could a discovery make the volunteer famous? Volunteers may also value being named on a NASA or ESA website more than (for example) being a co-author on an academic paper.

Ultimately, we would like all researchers working in the European Open Science Cloud to consider this daunting but thrilling question: what would you like a hundred thousand people to do for you? ∎

### About the Author

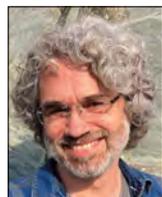

**Stephen Serjeant** is Professor of Astronomy at the Open University. Besides his work in star-forming galaxies, strong gravitational lensing and infrared astronomy, he has long-standing interests in open science, citizen science and machine learning. He wrote the advanced undergraduate / postgraduate textbook, Observational Cosmology, and has co-written several other books.

### Acknowledgements

SS has been supported by the ESCAPE project. ESCAPE - The European Science Cluster of Astronomy & Particle Physics ESFRI Research Infrastructures has received funding from the European Union's Horizon 2020 research and innovation programme under Grant Agreement no. 824064.

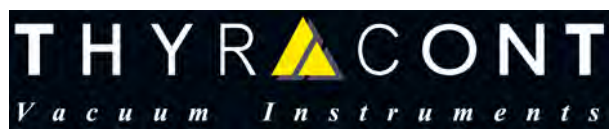

## VD800
## Compact vacuum meters.
## On the road to the future.

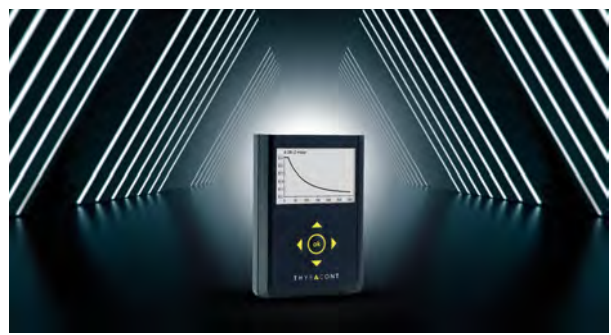

Clean design, intuitive menu navigation, large data logger and precise measurement results – with its new VD800 series Thyracont Vacuum Instruments is on the road to the future, consistently concentrating on the users' needs.

The VD800 vacuum meters measure absolute pressure in a large range of 2000 to 5e-5 mbar and additionally relative pressure in the range of -1060 to +1200 mbar. Their big graphic display shows current measurement values and pressure graphs as well as minimum and maximum pressure. The 4+1 membrane keypad provides a comfortable, menu-driven operation. An integrated data logger saves multiple measurement series with their RTC data. With sampling rates from 50 ms to 60 s the instruments are not only of interest for fast vacuum processes but also for long-term monitoring.

For charging the devices via USB-C interface, a standard power supply is sufficient. Also, the USB-C interface – or optional Bluetooth® LE for wireless data transfer – allows a direct read-out of the measured data and the export of measurement series stored in the data logger.

Mobile use or permanent installation: The VD800 vacuum meters can be connected to vacuum pumps or plants, linked to external sensors or directly placed into vacuum chambers. The compact VD800 series is perfect for service use, quality control and maintenance work or for leakage testing via rising-pressure method.

Thyracont will present the new VD800 series at Hannover Messe hall 13, booth C77.

**For more information, contact:**
Thyracont Vacuum Instruments
https://thyracont-vacuum.com/en
Phone: +49 851 95986-0